# Diffusion wave modes and oscillating motions in superfluid $He^3 - A$.

**Sh. E. Kekutia, N. D. Chkhaidze and D.G. Sanikidze**

Institute of Cybernatics, 5 Sandro Euli, Tbilisi, 0186, Georgia

E-mail: kekuka@yahoo.com; nikolozi1951@yahoo.com

**CONTENTS**

**Abstract**

Diffusion vibrational modes are studied in superfluid $He^3 - A$ in zero magnetic fields for different angles $\vartheta$ between the orbital axis $\vec{l}$ and the wave vector $\vec{q}$. The dispersion relation for these modes is found to depend on the wave polarization. It is shown that in addition to the normal component velocity, the superfluid component velocity also oscillates in the diffusion modes of $He^3 - A$. The frictional forces due to viscous waves in $He^3 - A$, exerted in the plane surface which is in contact with a superfluid liquid layer of finite thickness and performs a simple harmonic oscillatory motion, are calculated. We also consider, volume of $He^3 - A$ restricted by two parallel infinite surfaces, when the lower surface accomplishes simple harmonic oscillation and we calculate the frictional forces exerted on surfaces. It is found that the frictional force has not only parallel, but also a perpendicular component relative to the direction of oscillations.

## 1. Introduction.

Superfluid $He^3$ is a complex and beautiful system where the internal degrees of freedom play a prominent role, leading to a large array of phenomenon that is not seen in conventional $s$ wave superconductivity or in superfluid $He^4$. The complex phase diagram shows three different superfluid phases of liquid $He^3$, each with its own unique set of properties. Superfluid $He^3$ can serve as a model system for processes in the early universe. As the universe cooled, it made transitions from a featureless hot universe to the universe we know today via phase transitions involving broken symmetry. Studies of this exotic superfluid have stimulated investigators in a wide variety of other fields. It is well known how superfluid $He^3$ research can be linked to high $T_c$ superconductivity, heavy fermions superconductivity and liquid crystals in condensed matter physics. The superfluid phase transitions in $He^3$ may also serve as a model for transitions which occurred in the early universe.

So, during the intensive investigation of superfluid $He^3$ in the past decade [1,3], the idea of broken symmetry has once again proven to be a powerful and unifying tool in understanding condensed many-body systems. Below the superfluid transition, $He^3$ simultaneously breaks all the continuous symmetries that are broken separately in superfluid $He^4$, nematic liquid crystals, and antiferromagnets. Detailed study of superfluid $He^3$ has added a new facet to the concept of broken symmetry, contributing to a fuller appreciation of Landau's idea and enabling us to predict, and understand, collective behavior unsuspected until recently.

This new facet is the breaking of a relative symmetry. For example, the $He^3 - B$ phase breaks a relative spin-orbit symmetry [1] that is a linear combination between rotations in spin and orbital space. Its dynamics, therefore, include aspects of antiferromagnets [4] and of nematic liquid crystals.

The magnetic $A_1$-phase breaks also another relative symmetry, the spin-orbit-gauge symmetry. Corruccini and Osheroff investigated the Goldstone mode, a spin-temperature wave, of this broken symmetry [10]. This is the first time that any direct consequences of a broken relative symmetry have been detected. The Goldstone mode of $He^4$ superfluid is second sound, a temperature wave.

The $A$-phase of superfluid $He^3$ breaks the relative gauge-orbit symmetry. It is characterized by the equivalence between gauge transformation and a certain orbital rotation.

The superfluid $He^3$ revealed a new aspect in the concept of broken symmetry. That is, the gauge, spin, and orbital symmetries in superfluid $He^3$ are broken in certain combinations, and hence the typical properties of $HeII$ as an antiferromagnets and a liquid crystal are not manifested independently, but are mutually related. A peculiar broken of symmetry in the $A$-phase which preserves the hybrid orbital-gauge symmetry leads to a unique relation between the superfluid and liquid-crystal properties of the $A$-phase [1,2].

The present work aims to investigate the effect of these peculiarities on diffusion oscillatory modes in superfluid $He^3 - A$.

It is well known that, in addition to the weakly attenuating oscillating mode with the dispersion relation $\omega \propto q$, liquids exhibit strongly attenuating diffusion modes with $\omega \propto q^2$, where the proportionality factor includes some kinetic coefficients. In ordinary liquids, such waves include the viscous wave with the dispersion relation $i\omega = \left(\nu/\rho\right) q^2$ ($\nu$ is the viscosity and $\rho$ the density of the liquid), in which the only oscillating quantity is the velocity $\vec{\upsilon}$ perpendicular to the vector $\vec{q}$, and the temperature wave with the dispersion relation $i\omega = \chi q^2$ ($\chi$ is the thermal diffusivity of the medium). In superfluid $He^4$, the temperature oscillates in the form of a weakly attenuating mode, and the only diffusion mode is the viscous wave with the dispersion relation $i\omega = \nu q^2 / \rho_n$, where $\rho_n$ is the density of the normal component. In this wave, it is only the normal component velocity $\vec{\upsilon}^n$ perpendicular to the vector $\vec{q}$ that oscillates. As in an ordinary liquid, the viscous wave in superfluid $He^4$ is a transverse wave, dispersion relation of which does not depend on the polarization of waves.

The dispersion relation for viscous waves in an anisotropic medium (like a liquid crystal) depends on the direction of polarization. Besides, the existence of an additional dynamic variable for nematic liquid crystals, the director $\vec{n}$ specifying a preferred direction in space, leads to the emergence of new types of diffusive oscillating modes typical of these materials [3].

A similar situation is observed for anisotropic superfluid liquid $He^3 - A$ in which the preferred direction in space, i.e., the quantization axis for the orbital moment of Cooper pairs, is determined by the unit vector $\vec{l}$. For this reason, in addition, to the viscous wave in which the main oscillating quantity is $\vec{\upsilon}^n$, $He^3 - A$ acquires an additional viscous wave associated with the hydrodynamic variable $\vec{l}$ specifying the main oscillating quantity in this wave, i.e., $\vec{\upsilon}^l = \left(\hbar/2m\right) curl\, \vec{l}$ having the dimensions of velocity [4].

It can be easily seen that, in contrast to superfluid $He^4$, $He^3 - A$ has diffusion oscillating modes with an interesting peculiarity of another type: the superfluid component is also involved in the general case in oscillations occurring in waves in which $\vec{\upsilon}^n$ or $\vec{\upsilon}^l$ oscillates in the $\left(\vec{q}, \vec{l}\right)$ plane. Indeed, for sufficiently small values of $q$, the frequency of oscillations in viscous waves is $\omega \ll c_1 q$, $\omega \ll c_2 q$ ($c_1$ and $c_2$ are the velocities of first and second sound respectively), and hence we can neglect the density and temperature oscillations. In this case, it follows from the law of conversation of entropy of a liquid that $div \vec{v}^n = 0$, and hence $\vec{q} \cdot \vec{v}^n = 0$, for a plane wave, while from the continuity equations we have $div \vec{g} = 0$, and hence $\vec{q} \cdot \vec{g} = 0$ for the total mass flux. In other words, the waves, under consideration, are transverse relative to oscillations of the normal component velocity and the total mass flux. In $He^3 - A$, however, the direction of the normal component flux $\vec{g}^n$ does not coincide with the direction of the velocity $\vec{v}^n$ of normal component, i.e., the wave vector is not perpendicular to $\vec{g}^n$ when $\vec{v}^n$ is not directed along or at right angles to $\vec{l}$ in view of the tensor nature of the normal component density $\rho_{ik}^n$. Consequently, $\vec{q} \cdot \vec{g}^S \neq 0$ ($\vec{g}^S$ is the superfluid flux), and the superfluid component must participate in oscillations. The same is true for the mode in which $\vec{v}^l$ oscillates since $div \vec{v}^l = (\hbar/2m) div rot \vec{l} = 0$, and the coefficient $C_{ij}$ appearing with $\vec{v}^l$ in the expression for the total mass flux $\vec{g}$ is also a tensor.

In the present work, we shall analyze diffusion modes in $He^3 - A$ in zero magnetic field for an arbitrary angle $\vartheta$ between the orbital axis $\vec{l}$ and the wave vector $\vec{q}$ and calculate the frictional forces caused by these waves and acting in the plane surface which is in contact with the superfluid liquid layer of a finite thickness and performs a simple harmonic oscillatory motion in its plane.

## 2. Analysis of diffusion oscillatory modes

Let us consider an unbounded volume of $^3He - A$ in zero magnetic fields. In this case, the effect of mixing of spin and spatial variables is not observed. Considering small oscillations of spatial variables, we shall use the system of equations of orbital hydrodynamics of $^3He - A$ in linearized form [5]:

$$\dot{\rho} + \partial_i g_i = 0 \, ; \quad \dot{g}_i + \partial_i g_{ij} = 0 \, ; \quad \left( \dot{\rho s} \right) + \partial_i \left( \rho s v_i^n + \frac{q_i}{T} \right) = 0 \, ;$$

$$\frac{\hbar}{2m} \dot{\varphi} + J_\varphi = 0; \quad \frac{\hbar}{2m} \dot{l}_i + X_i = 0 \tag{1}$$

Here $g_i = \rho_{ij}^n \upsilon_j^n + \rho_{ij}^S \upsilon_j^S + C_{ij}\left(curl\,\vec{l}\right)_j$ is the total mass flux, where $\rho_{ij}^n = \rho_\parallel^n l_i l_j + \rho_\perp^n \delta_{ij}^T$ ($\delta_{ij}^T = \delta_{ij} - l_i l_j$) the density tensor of the normal component, and $\rho_{ij}^S = \rho_\parallel^S l_i l_j + \rho_\perp^S \delta_{ij}^T$, the density tensor of the superfluid component. Similarly, for the orbital component $C_{ij} = C_\parallel l_i l_j + C_\perp \delta_{ij}^T$,

$$\sigma_{ij} = P\delta_{ij} - \left(\alpha_1 \delta_{jk}^T l_i + \alpha_2 \delta_{ik}^T l_j\right)\Psi_k - \gamma' \varepsilon_{ijk} l_k \left(\nabla \lambda^S\right) +$$
$$+ \left[\gamma_{il}^{(1)} \varepsilon_{jpq} + \gamma_{jq}^{(2)} \varepsilon_{ipl} + \gamma_{jl}^{(3)} \varepsilon_{ipq} + \gamma_{iq}^{(3)} \varepsilon_{ipl}\right] l_p \partial_q \upsilon_l^n - \nu_{ijkl}\left(\partial_k \upsilon_l^n + \partial_l \upsilon_k^n\right) - \xi_{ij}\left(\nabla \lambda^S\right) + \xi_{ijk}\Psi_k$$

is the momentum flux tensor, where P is the pressure, $\gamma' = \hbar/2m$ (m is the mass of the $He^3$ atom), $\gamma_{il}^{(1)}$, $\gamma_{il}^{(2)}$ and $\gamma_{il}^3$ are the reactive coefficients of the form $\gamma_{ij} = \gamma_\parallel l_i l_j + \gamma_\perp \delta_{ij}^T$, $\xi_{ij}$ is the dissipative coefficient of the same form (it should be noted that $\gamma_\perp^{(1)} = \gamma_\perp^{(2)}$, $\gamma_\parallel^{(1)} = \gamma_\parallel^{(2)} = \gamma_\parallel^{(3)}$),

$$\nu_{ijkl} = \nu_1 l_i l_j l_k l_l + \nu_2 \delta_{ij}^T \delta_{kl}^T + \frac{\nu_3}{2}\left(\delta_{ij}^T l_k l_l + \delta_{kl}^T l_i l_j\right) + \frac{\nu_1}{2}\left(\delta_{ik}^T \delta_{jl}^T + \delta_{il}^T \delta_{jk}^T\right) + \frac{\nu}{2}\left(\delta_{ik}^T l_j l_l +\right.$$
$$\left. + \delta_{jl}^T l_i l_k + \delta_{il}^T l_j l_k + \delta_{lk}^T l_i l_l\right)$$

The normal viscosity tensor, $\xi_{ijk} = \xi\left(l_k \varepsilon_{ijp} + l_i \varepsilon_{kjp}\right)l_p$ ($\xi$ is the additional viscosity typical of $He^3 - A$),

$$\lambda_i^S = \rho_{ij}^S\left(\upsilon_j^S - \upsilon_j^n\right) + C_{ij}\left(rot\,\vec{l}\right)_i, \quad \Psi_i = -\frac{2m}{\hbar}\partial_i\left\{C_\perp\left(\upsilon_k^S - \upsilon_k^n\right)\varepsilon_{kji} + \left(C_\parallel - C_\perp\right)\left[\left(\vec{\upsilon^S} - \vec{\upsilon^n}\right)\vec{l}\right]\times\right.$$
$$\left.\times l_k \varepsilon_{kji} + K_1\left(\nabla \vec{l}\,\delta_{ij}\right) + K_2\left(\vec{l}\nabla\times\vec{l}\right)l_k \varepsilon_{kji} + K_3\left[l_j\left(\vec{l}\nabla\right)l_i - l_i\left(\vec{l}\nabla\right)l_j\right]\right\}$$

the thermodynamic force conjugates to $\nabla \vec{l}$, in which $K_1$, $K_2$, $K_3$ are the orbital rigidity coefficients, $s$ the entropy per unit mass, $q_i = k_{ij}\partial_j T$ the heat flux, where $k_{ij} = k_\parallel l_i l_j + k_\perp \delta_{ij}^T$ is the thermal conductivity tensor, and $T$ the temperature, $\varphi$ is the superfluid velocity potential, $\left[\vec{\upsilon^S} = \left(\hbar/2m\right)\nabla\varphi\right]$, $J_\varphi = \mu - \gamma'\vec{l}\left(\nabla\times\vec{\upsilon^n}\right) - \zeta\nabla\lambda^S - \xi_{ij}\partial_j \upsilon_i^n$, $\mu$ chemical potential, $\zeta$ the dissipation coefficient, $X_i = -\left(\alpha_1 \delta_{ij}^T l_k + \alpha_2 \delta_{ik}^T l_j\right)\partial_j \upsilon_k^n + \eta\delta_{ij}^T \Psi_j - \xi_{kji}\times\partial_k \upsilon_j^n$, $\eta$ is the Cross-Anderson orbital viscosity.

We choose the coordinate system so that $z \parallel \vec{l}$, and the y-axis is chosen in the $\left(\vec{q}, \vec{l}\right)$ plane. Further, we assume that all the variables vary according to the law $\exp i(q_z z + q_y y - \omega t)$, and the oscillation of $l_z$ can be neglected in the linear approximation.

Then Eqs. (1) are reduced to a cumbersome and complicated system of linear algebraic equations which will be solved under the assumption that the reactive and dissipative coefficients appearing in the hydrodynamic equations of the $A$-phase are connected through the following order-of-magnitude relations [6]:

$$\frac{\nu}{\rho} \gg \frac{\hbar}{2m} \sim \alpha \sim \xi \sim \eta\rho \sim \frac{C}{\rho} \sim \frac{\gamma_\parallel}{\rho} K\left(\frac{2m}{\hbar\rho}\right) \sim \zeta\rho \qquad (2)$$

Using these relations, we can single out (in the algebraic equations) the terms with small coefficients of oscillating quantities of the order of $\nu/\rho$ and $\hbar/2m$. The diffusion oscillatory modes will obviously belong to two essentially different types with the dispersion relations $i\omega \sim \nu q^2/\rho$ and $i\omega \sim (\hbar/2m) q^2$, in the first approximation. For the same value of $q$, the second frequency is small as compared to the first one, and hence the oscillations of the first type can be regarded as the fast and of the second type as slow as in the case of pneumatics [3].

The condition of existence of nontrivial solutions of a system of homogeneous linear algebraic equations is the equality to zero of the determinant formed by the coefficients of unknown quantities. Hence we obtain the equation specifying the dispersion relation for oscillatory modes. There are two fast and two slow diffusion modes. In all these modes, we can neglect the oscillations of pressure and temperature. Let, consider the case when $\vartheta$ is not close to $0, 180^0$ or $90^0$. In this case, $\upsilon_x^n$ oscillates in one of the fast modes. The dispersion relation for this mode is given by

$$i\omega = \frac{\nu_4 q_y^2 + \nu_5 q_z^2}{\rho_\perp^n} = \frac{\nu_4 \sin^2 \vartheta + \nu_5 \cos^2 \vartheta}{\rho_\perp^n} q^2.$$

The ratio of other oscillating quantities to $\upsilon_x^n$ is small:

$$\frac{\upsilon_y^n}{\upsilon_x^n} \sim \frac{\upsilon_z^n}{\upsilon_x^n} \sim \frac{\upsilon_x^l}{\upsilon_x^n} \sim \frac{\upsilon_y^l}{\upsilon_x^n} \sim \frac{\upsilon_z^l}{\upsilon_x^n} \sim \frac{\upsilon_y^S}{\upsilon_x^n} \sim \frac{\upsilon_z^S}{\upsilon_x^n} \sim \frac{\hbar}{2m}\frac{\rho}{\nu} \ll 1.$$

For the dispersion relation of the second fast mode, we have

$$i\omega = \frac{\left\{\left(\rho_\perp^S \sin^2\vartheta + \rho_\parallel^S \cos^2\vartheta\right)\left[v_5 \cos^4\vartheta + 2(v_1+v_2+v_4-v_3-v_5)\times\cos^2\vartheta\sin^2\vartheta + v_5 \sin^4\vartheta\right]\right\}}{\left\{\left(\rho_\perp^S \sin^2\vartheta + \rho_\parallel^S \cos^2\vartheta\right)\left(\rho_\perp^n \cos^2\vartheta + \rho_\parallel^n \sin^2\vartheta\right) + \left(\rho_\parallel^S - \rho_\perp^S\right)\left(\rho_\parallel^n - \rho_\perp^n\right)\cos^2\vartheta\sin^2\vartheta\right\}} q^2$$

The main oscillating quantities in this mode are $v_y^n, v_z^n, v_y^S$ and $v_z^S$, while the oscillations of the remaining quantities can be neglected. For a given law of variation of $v_z^n$, we obtain the laws of variation of the remaining quantities:

$$\frac{v_z^S}{v_z^n} = \frac{\left(\rho_\perp^n - \rho_\parallel^n\right)\cos^2\vartheta}{\rho_\perp^n \sin^2\vartheta + \rho_\parallel^S \cos^2\vartheta}, \quad \frac{v_y^S}{v_z^n} = \frac{\left(\rho_\perp^n - \rho_\parallel^n\right)\sin\vartheta\cos\vartheta}{\rho_\perp^n \sin^2\vartheta + \rho_\parallel^S \cos^2\vartheta},$$

$$\frac{v_y^n}{v_z^n} = -ctg\,\vartheta, \quad \frac{v_x^n}{v_z^n} \sim \frac{v_x^l}{v_z^n} \sim \frac{v_y^l}{v_z^n} \sim \frac{v_z^l}{v_z^n} \sim \frac{\hbar}{2m}\frac{\rho}{v} \ll 1.$$

The dispersion relation of the slow mode with the main oscillating quantity $v_x^l$ has the form

$$i\omega = \left(\frac{2m}{\hbar}\right)^2 \eta\left(K_1 q_y^2 + K_3 q_z^2\right),$$

and the ratio of $v_y^l$, $v_z^l$, $v_x^n$, $v_y^n$, $v_z^n$, $v_y^S$, $v_z^S$ to $v_x^l$ is the order of $\left(\hbar\rho/2mv\right) \ll 1$.

The second slow mode is formed by a wave with the dispersion relation

$$i\omega = \left(\frac{2m}{\hbar}\right)^2 \eta\left[K_2 q_y^2 + K_3 q_z^2 - \frac{(C_\perp - C_\parallel)^2 q_y^2 q_z^2}{\rho_\perp^S q_y^2 + \rho_\parallel^S q_z^2}\right] \quad (3)$$

The quantities $v_y^l$, $v_z^l$, $v_y^S$, $v_z^S$ have the same order of magnitude, and the ratios of $v_x^n$, $v_y^n$, $v_z^n$, to these quantities are of the order of $\left(\hbar\rho/2mv\right) \ll 1$.

For the ratio of the main oscillating quantities accompanying mode (3), we have

$$\frac{v_y^S}{v_z^l} = \frac{\hbar^{-1}(C_\perp - C_\parallel) q_y q_z}{\rho_\perp^S q_y^2 + \rho_\parallel^S q_z^2}, \quad \frac{v_z^S}{v_z^l} = \frac{\hbar^{-1}(C_\perp - C_\parallel) q_z^2}{\rho_\perp^S q_y^2 + \rho_\parallel^S q_z^2}, \quad \frac{v_y^l}{v_x^l} = -\frac{q_z}{q_y}$$

It should be noted that all the modes listed above are transverse relative to the oscillations of the normal velocity $\vec{v}^n$ and rotational velocity $\vec{v}^l$ are longitudinal relative to the oscillations of the superfluid velocity $\vec{v}^S$.

Let us consider the results of calculations for $\vartheta = 90^0$ $(q_z = 0)$ and $\vartheta = 0, 180^0$ $(q_y = 0)$. In this case, the effect of participation of the superfluid component in diffusion modes is either absent or very weak.

If $\vartheta = 90^0$, we get for fast oscillatory modes

$$i\omega = \frac{v_5 q^2}{\rho_\parallel^n}, \quad i\omega = \frac{v_4 q^2}{\rho_\perp^n} \tag{4}$$

In the first mode, $v_z^n$ oscillates, the oscillation of $\vec{v}^l$ is small, while pressure, temperature, superfluid velocity and other components of the normal velocity do not oscillate at all. In the second mode of (4), $v_x^n$ oscillates, the oscillations of pressure, temperature and superfluid velocity are small, while the orbital vector (and hence $\vec{v}^l$) does not participate in the oscillatory process.

For slow modes, we can write

$$i\omega = \left(\frac{2m}{\hbar}\right)^2 \eta K_1 q^2 \tag{5}$$

$$i\omega = \left(\frac{2m}{\hbar}\right)^2 \eta K_2 q^2 \tag{6}$$

In these modes, the oscillating quantities are $\vec{v}^l$ and $v_x^n$. In the first mode (5), the main oscillating quantity is $v_x^l$, while the oscillations of $v_y^l$, $v_z^l$ and $v_z^n$ are small. In the second mode (6), the variations of $v_y^l$ and $v_z^l$ are of the same order, while $v_x^l$ and $v_z^n$ they are small.

In our earlier work [7], the case of $\vartheta = 0$ and $\vartheta = 180^0$, is considered and an interesting result is obtained for diffusion modes which are in the case the circular polarization waves with different dispersion relations for waves with right and left polarization. In two of these modes, the quantities $v_+^n = v_y^n + iv_x^n$ and $v_+^l = v_y^l + iv_x^l$, oscillate, while $v_-^n = v_y^n - iv_x^n = 0$ and $v_-^l = v_y^l - iv_x^l = 0$. We shall call them the "plus" modes. In the remaining two modes, $v_-^n$ and $v_-^l$ oscillate, while $v_+^n$ and $v_+^l$ are equal to zero. We shall call them the "minus" modes. The dispersion relations for these modes

$$q_\pm^2 = \frac{i\omega}{D_0}\left(1 - \frac{D_4 + R_4 D_3}{D_0} \mp i\frac{R_1 + R_3 R_4}{D_0}\right); \tag{7}$$

$$\alpha_\pm^2 = \frac{i\omega}{D_4}\left(1 + \frac{R_2 D_3 D_4 + D_2 D_3 - R_2 R_3}{D_0 D_4} \pm i\frac{R_3 R_4 D_4 + R_3 D_2 + R_2 D_3}{D_0 D_4}\right). \tag{8}$$

Here, we have introduced the following notation:

$$R_1 = \frac{\gamma_\parallel^{(2)} + 2\alpha_2 m C_\perp/\hbar}{\rho_\perp^n}, \quad R_2 = \frac{2m\alpha_2 K_2}{\hbar\rho_\perp^n}, \quad R_3 = \frac{2m\alpha_2}{\hbar}, \quad R_4 = \frac{C_\perp}{\rho_\perp^n}, \quad D_0 = \frac{v_5}{\rho_\perp^n},$$

$$D_2 = \frac{2mK_3}{\hbar\rho_\perp^n}, \quad D_4 = \eta\left(\frac{2m}{\hbar}\right)^2 C_\perp + \frac{2m\xi}{\hbar}, \quad D_4 = \left(\frac{2m}{\hbar}\right)^2 \eta K_3, \tag{9}$$

$q$ being the wave vector of fast modes (i.e., viscous waves with a large penetration depth) and $\alpha$ the wave vector of slow modes (i.e., viscous waves with a smaller penetration depth).

The ratio of oscillating quantities is given by

$$b_+ = \left(\frac{v_+^l}{v_+^n}\right)_{q_+} = \frac{\hbar}{2m}\frac{(-R_3 + iR_3)q_+^2}{\omega + iD_4 q_+^2}, \quad a_+ = \left(\frac{v_+^l}{v_+^n}\right)_{\alpha_+} = \frac{\hbar}{2m}\frac{(-R_3 + iD_3)\alpha_+^2}{\omega + iD_4 \alpha_+^2},$$

$$b_- = \left(\frac{v_-^l}{v_-^n}\right)_{q_-} = \frac{\hbar}{2m}\frac{(R_3 + iD_3)q_-^2}{\omega + iD_4 q_-^2}, \quad a_- = \left(\frac{v_-^l}{v_-^n}\right)_{\alpha_-} = \frac{\hbar}{2m}\frac{(R_3 + iD_3)\alpha_-^2}{\omega + iD_4 \alpha_-^2}. \tag{10}$$

Using estimations (2), we can easily see that $\left(v^l/v^n\right) \approx \left(\hbar\rho/2mv\right) \ll 1$, for fast modes and $\left(v^n/v^l\right) \approx \left(\hbar\rho/2mv\right) \ll 1$ for slow modes.

It should be noted that the results obtained in Ref.7 are presented here in a somewhat different form which will be used in the next section. The waves with the right and left circular polarizations are different since the opposite direction $\vec{l}$ and $-\vec{l}$ in $^3He - A$ are physically nonequivalent. In this respect, $^3He - A$ differs from pneumatics in which the directions $\vec{n}$ and $-\vec{n}$ of the director are indistinguishable, and a similar effect is not observed.

### 3. Vibrating plane covered by a layer of superfluid $^3He - A$

Processes occurring in superfluid liquids with confining geometrical conditions have been become recently an object of thorough investigations. In Ref.8, the effect of boundaries on the $^3He$ state in films having a thickness of the order of correlation length was studied theoretically, and new properties were predicted both for the $B-$ and $A-$ phase. They differ in their physical properties from bulk phases.

It is well known that the $^3He - A$ phase borders the normal $^3He$ phase in zero magnetic field under a pressure $P > P_0 \approx 21$ bar, while the $^3He - B$ phase is stable at $P < P_0$. However, in a sufficiently thin film of superfluid $^3He$ deposited on a solid substrate, the $A-$ phase is found to be more stable than the $B-$ phase even under the saturated vapor pressure. This is due to the fact that the confining surface (a solid or a free surface) does not change the order parameter of the $A-$ phase in view of the inherent preferred direction of the orbital axis $\vec{l}$, and hence its rotation towards the normal to the surface does not lead to a loss in the condensation energy. In contrast to the $A-$ phase, the order parameter of the $B-$ phase undergoes considerable changes due to the emergence of a preferred direction (which does not exist in the bulk) due to the tendency of orbital moments to arrange themselves along the normal to the surface. The possibility of a transition to the planar phase is ruled out by the presence of paramagnon effects. It was shown in Ref. 9 that the $A-$ phase is stabilized for films of thickness $d < d_c(T) = d_0\left(1 - T/T_c\right)^{-1/2}$, where $d_0 \approx 15\xi_0$ ($\xi_0$ is the coherence length), while the $B-$ phase is stable for $d > d_c$. For a given thickness $d > d_0$ of the film, both $A$ and $B$ phases can exist. A transition from the $A-$ to $B-$ phase takes place at a temperature $T_{AB} = \left[1 - \left(d_0/d\right)^2\right]T_c$.

Using the solutions (7) and (8) of the wave equation obtained above, we can determine the frictional forces exerted on the solid surface vibrating along the $y$-axis with a velocity $u = u_0 \exp(-i\omega t)$ and covered with a layer of superfluid $^3He - A$ of thickness $d$, the upper surface of the liquid is free. For an ordinary viscous fluid [9] and for superfluid $^4He$, a frictional force emerges only in the direction of the plane vibrations. It was found that the frictional force emerging in superfluid $^3He-A$ has components directed both parallel and perpendicular to the direction of plane vibrations. Since the orbital moment of a Cooper pair tends to align itself at right angles to the walls, we shall assume that in the given geometry the vector $\vec{l}$ in equilibrium is directed along the $z-$ axis. We consider the motion emerging in the liquid as the superposition of viscous waves excited near the solid plane and reflected by the free surface. Then we can write the following expressions for oscillating quantities:

$$\upsilon_+^n = \left(c_1 e^{iq_+ z} + c_1' e^{-iq_+ z} + c_2 e^{i\alpha_+ z} + c_2' e^{-i\alpha_+ z}\right) e^{-i\omega t}$$

$$\upsilon_-^n = \left(c_3 e^{iq_- z} + c_3' e^{-iq_- z} + c_4 e^{i\alpha_- z} + c_4' e^{-i\alpha_- z}\right) e^{-i\omega t} \quad (11)$$

$$l_+ = \left[\frac{(-R_3 + iD_3)q_+}{\omega + iD_4 q_+^2}\left(c_1 e^{iq_+ z} - c_1' e^{-iq_+ z}\right) + \frac{(-R_3 + iD_3)\alpha_+}{\omega + iD_4 \alpha_+^2}\left(c_3 e^{i\alpha_+ z} + c_3' e^{-i\alpha_+ z}\right)\right] e^{-i\omega t}$$

$$l_- = \left[\frac{(-R_3 + iD_3)q_-}{\omega + iD_4 q_-^2}\left(c_1 e^{iq_- z} - c_1' e^{-iq_- z}\right) + \frac{(-R_3 + iD_3)\alpha_-}{\omega + iD_4 \alpha_-^2}\left(c_3 e^{i\alpha_- z} + c_3' e^{-i\alpha_- z}\right)\right] e^{-i\omega t}$$

While writing the expressions for the quantities $l_+ = l_y + il_x$ and $l_- = l_y - il_x$, we have used expressions (10) and the well-known relations

$$\upsilon_+^l = -\frac{\hbar}{2m}\frac{\partial l_+}{\partial z}, \quad \upsilon_-^l = -\frac{\hbar}{2m}\frac{\partial l_-}{\partial z}$$

The boundary conditions have the form

$$l_+ = l_- = 0, \quad \upsilon_x^n = 0, \quad \upsilon_y^n = u = u_0 e^{-i\omega t} \text{ for } z = 0, \quad (12)$$

$$l_+ = l_- = 0, \quad \sigma_+ = \sigma_{yz} + i\sigma_{xz} = 0, \quad \sigma_- = \sigma_{yz} - i\sigma_{xz} = 0 \text{ for } z = d$$

The viscous stress tensor component $\sigma_{yz}$ and $\sigma_{xz}$ are equal to zero for $z = d$, since there is no viscous force on the free surface of the liquid. The quantities $\sigma_+$ and $\sigma_-$ are defined as

$$\sigma_+ = \left\{\left[R_1 - iD_1 - \frac{(R_2 - iD_2)(-R_3 + iD_3)q_+^2}{\omega + iD_4 q_+^2}\right]q_+ \left(c_1 e^{iq_+ z} - c_1' e^{-iq_+ z}\right) + \right.$$

$$\left. + \left[R_1 - iD_1 - \frac{(R_2 - iD_2)(-R_3 + iD_3)\alpha_+^2}{\omega + iD_4 \alpha_+^2}\right]\alpha_+ \left(c_2 e^{i\alpha_+ z} - c_2' e^{-i\alpha_+ z}\right)\right\}\rho_\perp^n e^{-i\omega t},$$

$$\sigma_- = \left\{ \left[ -R_1 - iD_1 + \frac{(R_2 + iD_2)(R_3 + iD_3)q_-^2}{\omega + iD_4 q_+^2} \right] q_- \left( c_3 e^{iq_- z} - c_3' e^{-iq_- z} \right) + \right.$$

$$\left. + \left[ -R_1 - iD_1 + \frac{(R_2 + iD_2)(R_3 + iD_3)\alpha_-^2}{\omega + iD_4 \alpha_-^2} \right] \alpha_- \left( c_4 e^{i\alpha_+ z} - c_4' e^{-i\alpha_+ z} \right) \right\} \rho_\perp^n e^{-i\omega t},$$

$$D_1 = D_0 + \frac{2m\xi C}{\hbar \rho_\perp^n}$$

The frictional force exerted per unit surface area is determined by the stress tensor components:

$$F_y = -\sigma_{yz} = -\frac{\sigma_+ + \sigma_-}{2}, \quad F_z = -\sigma_{xz} = \frac{\sigma_- - \sigma_+}{2} \tag{14}$$

Substituting expressions (11) and (13) for oscillating quantities into the boundary conditions (12), we obtain the system of equations for the amplitudes $c_i$ and $c_i'$ (i=1, 2, 3, 4) of the excited waves. Using them in (14), we get

$$F_y = \rho_\perp^n D_o q \, tg(qd) u, \quad F_x = \rho_\perp^n u \left[ \left( R_1 - \frac{R_2 D_3 + D_2 R_3}{D_4} \right) q \, tg\, qd - \frac{iD_o}{2}(q_+ tg\, q_+ d - q_- tg\, q_- d) \right] \tag{15}$$

In the limit $|q|d \ll 1$, using (7) and (8) we obtain from (15)

$$F_y = \rho_\perp^n D_o q^2 d\, u; \quad F_x = -\rho_\perp^n u q^2 d \frac{R_2 D_3 + D_2 R_3 + R_3 R_4 D_4}{D_4} \tag{16}$$

A Physical meaning can be attached only to the real parts of these complex expressions. Proceeding from (9) and (16), we can show

$$F_y = \omega \rho_\perp^n d\, u_0 \cos\left( \omega t - \frac{\pi}{2} \right); \tag{17}$$

$$F_x = \frac{2\alpha_2}{\eta} \left( \frac{2m}{\hbar} \eta C_\perp + \xi \right) \frac{\omega \rho_\perp^n}{v_5} d\, u_o \cos\left( \omega t + \frac{\pi}{2} \right) \tag{18}$$

These expressions show that the component (18) of the viscous force perpendicular to the direction of vibrations of the confining plane is much smaller than its longitudinal component (17).

## 4. Generating of frictional forces in the two parallel plates

## Saturated with superfluid $^3He - A$.

We use obtained wave solution to determine frictional forces, acting on each from two parallel, solid plates saturated with superfluid $^3He - A$, when one of the plates performs oscillating motion in its surface. In the case of usual liquid there generates frictional force only in direction of oscillation of the surface [3]. Analogous picture takes place in case of superfluid $^4He$. It seems that in superfluid $^3He - A$ emerging frictional force also has component, directed perpendicular as well as parallel to the direction of surface oscillation. The same effect was obtained in case above.

Let consider, volume $^3He - A$, restricted by two parallel infinite surfaces, distance (space) between that is defined with $d$. Suppose the lower surface accomplish simple harmonic oscillation across the axis $y$ with velocity $u = u_0 \exp(-i\omega t)$. We shall consider the upper surface rigid fixed (fastened). As orbital moment of Cooper pairs tries to be oriented perpendicular to walls, we shall consider that in given geometry the equilibrium $\vec{l}$ is directed across the axis $z$.

We are looking for generating motion in liquid as a superposition of shear modes along positive and negative directions of axis $z$. Then for oscillating quantities we have:

$$\upsilon_x^n = \left[ c_1 e^{i\alpha_+ z} + c_1^{'} e^{-i\alpha_+ z} + c_2 e^{i\alpha_- z} + c_2^{'} e^{-i\alpha_- z} + c_3 e^{iq_+ z} + c_3^{'} e^{-iq_+ z} + c_4 e^{iq_- z} + c_4^{'} e^{-iq_- z} \right] e^{-i\omega t}$$

$$l_x = \left[ a_+ c_1 e^{i\alpha_+ z} - a_+ c_1^{'} e^{-i\alpha_+ z} + a_- c_2 e^{i\alpha_- z} - a_- c_2^{'} e^{-i\alpha_- z} + b_+ c_3 e^{iq_+ z} + b_+ c_3^{'} e^{-iq_+ z} + b_- c_4 e^{iq_- z} - b_- c_4^{'} e^{-iq_- z} \right] e^{-i\omega t}$$

$$\upsilon_y^n = \left[ ic_1 e^{i\alpha_+ z} + ic_1^{'} e^{-i\alpha_+ z} - ic_2 e^{i\alpha_- z} - ic_2^{'} e^{-i\alpha_- z} + ic_3 e^{iq_+ z} + ic_3^{'} e^{-iq_+ z} - ic_4 e^{iq_- z} - ic_4^{'} e^{-iq_- z} \right] e^{-i\omega t} \quad (19)$$

$$l_y = \left[ ia_+ c_1 e^{i\alpha_+ z} - ia_+ c_1^{'} e^{-i\alpha_+ z} - ia_- c_2 e^{i\alpha_- z} + ia_- c_2^{'} e^{-i\alpha_- z} + ib_+ c_3 e^{iq_+ z} - \right.$$
$$\left. - ib_+ c_3^{'} e^{-iq_+ z} - ib_- c_4 e^{iq_- z} + b_- c_4^{'} e^{-iq_- z} \right] e^{-i\omega t}$$

Writing the relations (19) we utilize the conditions of polarization: $\upsilon_y^n = i\upsilon_x^n$, $l_y = il_x$ for right polarization and $\upsilon_y^n = -i\upsilon_x^n$, $l_y = -il_x$ for left polarization. The boundary conditions have the following form:

$$l_x = l_y = 0, \quad \upsilon_x^n = 0, \quad \upsilon_y^n = u = u_0 e^{-i\omega t} \quad \text{for } z=0$$

$$\upsilon_x^n = \upsilon_y^n = 0, \quad l_x = l_y = 0 \qquad \text{for z=d} \qquad (20)$$

Substituting expressions (19) for oscillating quantities into the boundary conditions (20), we obtain the system of equations for the amplitudes $c_i$ and $c_i^{'}$ of the excited waves:

$$c_1 + c_1^{'} + c_3 + c_3^{'} = u_0/2i; \quad c_1 e^{i\alpha_+ d} + c_1^{'} e^{-i\alpha_+ d} + c_3 e^{iq_+ d} + c_3^{'} e^{-iq_+ d} = 0;$$

$$a_+(c_1 - c_1^{'}) + b_+(c_3 - c_3^{'}) = 0; \quad a_+(c_1 e^{i\alpha_+ d} - c_1^{'} e^{-i\alpha_+ d}) + b_+(e^{iq_+ d} c_3 - c_3^{'} e^{-iq_+ d}) = 0;$$

$$c_2 + c_2^{'} + c_4 + c_4^{'} = -u_0/2i; \quad c_2 e^{i\alpha_- d} + c_2^{'} e^{-i\alpha_- d} + c_4 e^{iq_- d} + c_4^{'} e^{-iq_- d} = 0; \qquad (21)$$

$$a_-(c_2 - c_2^{'}) + b_-(c_4 - c_4^{'}) = 0; \quad a_-(c_2 e^{i\alpha_- d} - c_2^{'} e^{-i\alpha_- d}) + b_-(c_4 e^{iq_- d} - c_4^{'} e^{-iq_- d}) = 0;$$

Further it will be clear that, we need only some combinations of coefficients $c_i$ and $c_i^{'}$. So, full solution of system (21) due to its being huge we do not bring. The frictional force exerted per unit surface area of upper and lower surface is determined by the stress tensor components:

$F_y = -\sigma_{yz}$, $F_x = -\sigma_{xz}$ on the lower boundary, and $F_{y1} = \sigma_{yz}$, $F_{x1} = \sigma_{xz}$ on the motionless surface.

$$F_y = -v_5 q u\, ctg qd, \quad F_x = -(R_1 + R_3 R_4 + A + B\, tg qd) q u\, ctg qd$$

$$F_{y1} = \frac{v_5 q u}{\sin qd}; \quad F_{x1} = (R_1 + R_3 R_4 + A + B^{'} \sin qd) \left(\frac{q u}{\sin qd}\right)$$

where there are introduced the following notations:

$$A = \frac{iD_0}{2}\left(\frac{b_-}{a_-} - \frac{b_+}{a_+}\right)\frac{tg(d\alpha/2)}{tg(dq/2)}; \quad B = \frac{iD_0}{2q}(q_- ctg dq_- - q_+ ctg dq_+);$$

$$B^{'} = \frac{iD_0}{2q}\left(\frac{q_-}{\sin dq_-} - \frac{q_+}{\sin dq_+}\right)\sin dq.$$

Our results, corresponding to half-infinite volume in the limit $|q|d \gg 1$, coincide with expressions [6]. We have to notice, that contribution of $A$ coefficient must be considered when $|\alpha|d \leq 1$, and contribution due to $B, B'$-when $|q|d \geq 1$. In the limit cases the expressions of shear viscous force are simplified and get the following form:

$$F_x = -F_{x1} = \begin{cases} -\rho_\perp^n D_4^{-1}(R_1 D_4 - R_3 D_2 - R_2 D_3)(u/d), & d|\alpha| \ll 1; d|q| \ll 1 \\ -\rho_\perp^n (R_1 + R_3 R_4)(u/d), & d|\alpha| \gg 1; d|q| \ll 1 \end{cases}$$

When $d|q| \gg 1$, $d|\alpha| \gg 1$ we derive $F_x = \frac{1}{2} i \rho_\perp^n (R_1 + R_3 R_4)$, $F_{x1} = 0$.

In conclusion, we have studied diffusion modes in superfluid $^3He - A$ in zero magnetic fields. The dispersion relation for these modes is found to depend on the wave polarization. It is shown that in addition to the normal component velocity, the superfluid component velocity also oscillates in the diffusion modes of $^3He - A$. The frictional forces due to viscous waves in $^3He - A$, exerted on the plane surface which is in contact with a superfluid liquid layer of finite thickness and on the one of the plane surfaces when $^3He - A$ saturate the space between two infinite planes are calculated. It is found that the frictional force has not only parallel, but also a perpendicular component relative to the plane of oscillation. We notice that the component of the viscous force perpendicular to the direction of oscillations of the plane is smaller than its longitudinal component.

## L I T E R A T U R E